\documentclass{eptcs}
\providecommand{\event}{HSB 2012} 
\usepackage{breakurl}             

\usepackage{graphicx,amsmath,amssymb,fixmath,color}

\newcommand{\R}{{\Bbb R}}
\newcommand{\Z}{{\Bbb Z}}
\newcommand{\N}{{\Bbb N}}

\newcommand{\D}[2]{ \ensuremath{ \frac{d #1 }{d #2 } }}

\newcommand{\vect}[1]{\ensuremath{ \mathbold #1 } }

\newtheorem{theorem}{Theorem}[section]
\newtheorem{lemma}[theorem]{Lemma}
\newtheorem{proposition}[theorem]{Proposition}

\newtheorem{definition}[theorem]{Definition}

\title{Hybrid models of the cell cycle molecular machinery}
\author{Vincent Noel
  \institute{IRMAR UMR 6625, University of Rennes 1, Rennes, France}
\and Dima Grigoriev
    \institute{CNRS, Math\'ematiques, Universit\'e de Lille, 59655, Villeneuve d'Ascq, France}
\and Sergei Vakulenko
   \institute{Saint Petersburg State University of Technology and Design, St.Petersburg, Russia}
\and Ovidiu Radulescu
  \institute{DIMNP UMR CNRS 5235, University of Montpellier 2, Montpellier, France}}

\def\titlerunning{Hybrid models of the cell cycle}
\def\authorrunning{V. Noel, D. Grigoriev, S. Vakulenko,  \& O. Radulescu}

\begin{document}
\maketitle





\begin{abstract}
Piecewise smooth hybrid systems, involving continuous and discrete variables, are suitable models for describing
the multiscale regulatory machinery of the biological cells. In hybrid models, the discrete variables can switch
on and off some molecular interactions, simulating cell progression through a series of functioning modes. The
advancement through the cell cycle is the archetype of such an organized sequence of events.
We present an approach, inspired from tropical geometry ideas, allowing to reduce, hybridize and analyse cell
cycle models consisting of polynomial or rational ordinary differential equations.
\end{abstract}

\section{Introduction}
Hybrid systems are widely used in automatic control theory to cope with
situations arising when a finite-state machine is coupled to mechanisms that
can be modeled by differential equations \cite{matveev2000qualitative}.
It is the case of robots, plant controllers, computer disk drives, automated highway systems,
flight control, etc. The general behavior of such systems is to pass from one type of smooth
dynamics (mode) described by one set of differential equations
to another smooth dynamics (mode) described by another set of differential equations.
The command of the modes can be performed by changing one or several discrete variables.
The mode change can be accompanied or not by jumps (discontinuities) of the trajectories.
Depending on how the discrete variables are changed, there may be several types of
hybrid systems: switched systems \cite{shorten2007stability}, multivalued differential
automata \cite{tavernini1987differential}, piecewise smooth
systems \cite{filippov1988differential}. Notice that in the last case, the mode changes
when the trajectory attains some smooth manifolds. In these examples, the changes of discrete
variables and the evolution of continuous variables are deterministic. The
class of hybrid systems can be extended by considering stochastic dynamics of both continuous
and discrete variables, leading to piece-wise deterministic processes, switched diffusions or diffusions with
jumps \cite{TSI,crudu2009hybrid,crudu2011convergence,singh2010stochastic,bortolussi2011hybrid}.
Hybrid, differential, or stochastic
Petri nets provide equivalent descriptions of the dynamics and were also used
in this context \cite{david2005discrete}.


The use of hybrid models in systems biology can be justified by the
temporal and spatial multi-scaleness of biological processes, and by the
need to combine qualitative and quantitative approaches to study dynamics
of cell regulatory networks. Furthermore, hybrid modelling offers a good compromise
between realistic description of mechanisms of regulation and possibility of testing
the model in terms of state reachability and temporal logics
\cite{lincoln2004symbolic,mishra2009intelligently}. Threshold dynamics of gene
regulatory networks \cite{baldazzi2011qualitative,ropers2011model}
or of excitable signaling systems \cite{ye2008modelling}
has been modelled by piecewise-linear and piecewise-affine models.
These models have relatively simple structure and can, in certain cases, be
identified from data \cite{porreca2008structural,drulhe2008switching}.
Some methods were proposed for computing the set of reachable states of
piecewise affine models \cite{batt2008symbolic}.

Among the applications of hybrid modeling, one of the most important is the cell cycle regulation.
The machinery of the cell cycle, leading to cell division and proliferation, combines
slow growth, spatio-temporal re-organisation of the
cell, and rapid changes of regulatory proteins concentrations induced by post-translational
modifications. The advancement through the cell cycle is a well defined sequence of stages,
separated by checkpoint transitions.
This justifies hybrid modelling approaches, such as Tyson's  hybrid 
model of the mammalian cell cycle \cite{singhania2011hybrid}. This model is based on a
Boolean automaton whose discrete transitions trigger changes of kinetic parameters in
a set of ODEs.
The model has been used to reproduce flow cytometry data. Instead of building the hybrid model
from scratch, another strategy is to identify hybrid models from experimental or artificial
time series \cite{noel2010,noel2011,alfieri2011modeling}. The resulting cell cycle
hybrid models can be used for hypothesis testing, as they are
or as parts of larger, integrated models.

In this paper we develop ideas first introduced in \cite{SASB2011}. We discuss how a given model
of the cell cycle, based on ODEs, can be hybridized.
The hybridization, based on a tropical geometry
heuristics, unravels commonalities of cell cycle models. These systems
combine quasi-equilibrium states, represented by slow invariant manifolds and excitability, represented by rapid transitions to and from these manifolds.
With respect to \cite{SASB2011} we introduce several new
concepts and provide rigorous justification of the procedures. Two general hybridization procedures
called tropicalizations are introduced in section 2. The tropicalized dynamics is guaranteed to
be a good approximation for polynomial or rational systems with well separated terms and that
satisfy a condition called permanency. In subsection 2.2 we introduce the tropical equilibration
as a method to test permanency. In section 3 we apply these methods to a cell cycle biochemical
network model.

\section{Tropical geometry and hybridization}
\subsection{General settings}
In chemical kinetics, the reagent concentrations satisfy
ordinary differential equations:
\begin{equation}
\D{x_i}{t} = F_i (\vect{x}), \, 1 \leq i \leq n.
\label{mainsystem}
\end{equation}
Rather generally, the rates are rational functions of the concentrations and read
\begin{equation}
F_i (\vect{x})= P_i (\vect{x})/Q_i(\vect{x}),
\label{rational}
\end{equation}
where
$P_i(\vect{x}) = \sum_{\alpha \in A_i} a_{i,\alpha} \vect{x}^\alpha$,
$Q_i(\vect{x}) = \sum_{\beta \in B_i} b_{i,\beta} \vect{x}^\beta$,
are multivariate polynomials. Here
$\vect{x}^\alpha = x_1^{\alpha_1} x_2^{\alpha_2} \ldots x_n^{\alpha_n}$,
$\vect{x}^\beta = x_1^{\beta_1} x_2^{\beta_2} \ldots x_n^{\beta_n}$, $a_{i,\alpha}, b_{i,\beta}$, are
nonzero real numbers, and $A_i, B_i$ are finite subsets of $\N^n$.

Special case are represented by
\begin{equation}
F_i (\vect{x})= P_i^+(\vect{x}) - P_i^-(\vect{x}),
\label{laurent}
\end{equation}
where $P_i^+(\vect{x})$, $P_i^-(\vect{x})$ are Laurent polynomials with positive coefficients,
$P_i^{\pm}(\vect{x}) = \sum_{\alpha \in A_i^\pm} a_{i,\alpha}^\pm \vect{x}^\alpha$,
$a_{i,\alpha}^\pm > 0$, $A_i^\pm$ are finite subsets of $\Z^n$.

In multiscale biochemical systems, the various monomials of the Laurent polynomials have
different orders, and at a given time, there is only one or a few dominating terms.
Therefore, it could make sense to replace Laurent polynomials with positive real coefficients  $\sum_{\alpha \in A} a_{\alpha} \vect{x}^\alpha$, by max-plus polynomials
$max_{\alpha \in A} \{log( a_{\alpha}) + < log(\vect{x}), \alpha > \}$.

This heuristic can be used to associate a piecewise-smooth hybrid model
to the system of rational ODEs \eqref{mainsystem}, in two different ways.

The first method was proposed in \cite{SASB2011} and can be applied to any rational
ODE system defined by \eqref{mainsystem},\eqref{rational}:
\begin{definition}
We call complete tropicalization  of the smooth ODE
system \eqref{mainsystem},\eqref{rational} the following piecewise-smooth system:
\begin{equation}
\D{x_i}{t} = Dom P_i (\vect{x}) / Dom Q_i(\vect{x}),
\label{tcomplete}
\end{equation}
\noindent where $Dom \{a_{i,\alpha}  \vect{x}^\alpha \}_{\alpha \in A_i} =
sign(a_{i,\alpha_{max}}) exp [max_{\alpha \in A_i} \{ log( |a_{i,\alpha}|) + < \vect{u} , \alpha > \}]$.
$\vect{u} = (log x_1,\ldots,log x_n)$, and
$a_{i,\alpha_{max}},\, \alpha_{max}\in A_i$ denote the coefficient
of the monomial for which the maximum is attained. In simple words, $Dom$ renders the monomial
of largest absolute value, with its sign.
\end{definition}
The second method,proposed in \cite{savageau2009phenotypes}, applies to the systems \eqref{mainsystem},\eqref{laurent}.
\begin{definition}
We call two terms tropicalization  of the smooth ODE
system \eqref{mainsystem},\eqref{laurent} the following piecewise-smooth system:
\begin{equation}
\D{x_i}{t} = Dom P_i^+ (\vect{x}) - Dom P_i^- (\vect{x}),
\label{2terms}
\end{equation}
\end{definition}
The two-terms tropicalization was used in \cite{savageau2009phenotypes} to analyse the
dependence of steady states on the model parameters. The complete tropicalization
was used for the study of the model dynamics and for the model reduction \cite{SASB2011}.

For both tropicalization methods, for each occurrence of the Dom operator, one can introduce
a tropical manifold, defined as the subset of $\R^n$ where the maximum in Dom is attained by at least two
terms. For instance, for $n=2$, such tropical manifold is made of points, segments connecting these points, and half-lines. The tropical manifolds in such an arrangement decompose the space into sectors, inside which one monomial
dominates all the others in the definition of the reagent rates.
The study of this arrangement give hints on the possible steady states and attractors, as well as on their bifurcations.




\subsection{Justification of the tropicalization and some estimates}
In the general case, the tropicalization heuristic is difficult to justify by rigorous estimates,
however, this is possible in some cases. We state here some results in this direction.
To simplify, let us consider the class of polynomial systems, corresponding to mass action law chemical kinetics:
\begin{equation}
\frac{dx_i}{dt} = P_i(\vect{x}, \epsilon)=\sum_{j=1}^M M_{ij}(\vect{x},\epsilon), \quad M_{ij}=P_{ij}(\epsilon) \vect{x}^{\alpha_{ij}}
\label{chem}
\end{equation}
\noindent
where $\alpha_{ij}$ are multi-indices,
and $\epsilon$ is a small parameter. So, the right hand side of (\ref{chem})
is a sum of monomials. We suppose that coefficients $P_{ij}$ have different orders in $\epsilon$:
\begin{equation}
P_{ij}(\epsilon)=\epsilon^{b_{ij}} \bar P_{ij},
\label{eps}
\end{equation}
where $b_{ij} \neq b_{i'j'}$ for $(i,j)\neq (i',j')$.

We also suppose that the cone ${\bf R}_{> }=\{x: x_i  \ge 0 \}$ is invariant under dynamics (\ref{chem})
and initial data are positive:
$$
x_i(0) > \delta > 0.
$$
The terms (\ref{eps}) can have different signs, the ones with $\bar P_{ij} >0$ are production terms,
and those with $\bar P_{ij} <0$ are degradation terms.

From the biochemical point of view, the choice (\ref{eps}) is justified by the
fact that biochemical processes have many, well separated timescales.
Furthermore, we are interested in biochemical circuits that can  function
in a stable way.
More precisely, we use the permanency concept, borrowed from ecology
(the Lotka -Volterra model, see for instance \cite{y1996global}).
\begin{definition}
The system \eqref{chem} is permanent, if there are two constants $C_{-} >0$ and $C_{+} > 0$, and
 a function $T_0$, such that
\begin{equation}
 C_{-} < x_i(t) < C_{+},   \quad for \ all \ t > T_0 ( x(0)) \ and \ for \ every \ i.
\label{perm}
\end{equation}
We assume that $C_{\pm}$ and $T_0$ are uniform in (do not depend on) $\epsilon$ as $\epsilon \to 0$.
\end{definition}
For permanent systems, we can obtain some results justifying the two
procedures of tropicalization.
\begin{proposition}
\label{comparison}
Assume that system \eqref{chem} is permanent. Let $x$, $\hat x$ be the solutions to the Cauchy problem for (\ref{chem}) and (\ref{tcomplete}) (or (\ref{2terms})), respectively, with the same initial
data:
$$
  x(0)=\hat x(0).
$$
Then the difference $y(t)=x(t) - \hat x(t)$ satisfies  the estimate
\begin{equation}
     | y(t) |  < C_1 \epsilon^{\gamma} \exp(bt), \quad \gamma > 0,
\label{estdif}
\end{equation}
 where the positive constants $C_1, b$ are uniform in $\epsilon$.
If the original system (\ref{chem}) is structurally stable in the domain
$\Omega_{C_{-}, C_{+}}=\{x: C_{-} < |x| < C_{+} \}$,
then the corresponding tropical systems (\ref{tcomplete}) and (\ref{2terms})
are also permanent and there is an orbital topological equivalence
$\bar x = h_{\epsilon}(x)$ between the trajectories $x(t)$ and $\bar x(t)$
of the corresponding Cauchy problems. The homeomorphism $h_{\epsilon}$ is close to the identity as $\epsilon \to 0$.
\end{proposition}
The proof of the estimate (\ref{estdif}) follows immediately by the Gronwall lemma.
The second assertion follows directly
from the definition of  structural stability.

Permanency property is not easy to check. In the case of systems \eqref{chem}
we can make  a renormalization
\begin{equation}
  x_i=\epsilon^{a_i} \bar x_i
\label{renorm}
\end{equation}
and suppose that (\ref{perm}) holds for $\bar x_i$ with $C_i^{\pm}$ uniform in  $\epsilon$.


We seek for renormalization exponents $a_i$ such that only a few terms dominate all the others,
for each $i$-th equation \eqref{chem} as $\epsilon \to 0$.  Let us denote the number of terms with minimum
degree in $\epsilon$ for $i$-th equation as $m_i$. Naturally, $1 \le m_i \le M_i$.
After renormalization, we remove all small terms that have smaller orders in $\epsilon$  as $\epsilon \to 0$. We can call this procedure {\em tropical removing}. The system obtained can be named {\em tropically truncated system}.

Let us denote $\alpha_l^{ij}$ the $l^{th}$ coefficient of the multi-index $\vec{\alpha_{ij}}$.
If all $m_i=1$ then we have the following truncated system
\begin{equation}
\frac{d\bar x_i}{dt} =\epsilon^{\mu_i} F_{i}(\bar {\bf x}), \quad F_{i}=p_{ij(i)} \bar {\bf x}^{\alpha^{ij(i)}},
\label{chemshort}
\end{equation}
where
\begin{equation}
\mu_i=\gamma_{i j(i)} \sum_{l=1}^n \alpha_l^{ij(i)} a_l
\label{mu1a}
\end{equation}
and
\begin{equation}
\mu_i > \gamma_{i j} \sum_{l=1}^n \alpha_l^{ij} a_l \quad for \ all \  j \ne j(i).
\label{mu2}
\end{equation}
 If all $m_i=2$, in order to find possible renormalization exponents $a_i$,  it is necessary to resolve a family of linear programming problem. Each problem is defined by a set of pairs $(j(i), k(i))$ such that $j(i) \ne k(i)$. We define $\mu_i$ by
\begin{equation}
\mu_i=\gamma_{i j(i)} + \sum_{l=1}^n \alpha_l^{ij(i)} a_l=\gamma_{i k(i)} + \sum_{l=1}^n \alpha_l^{ik(i)} a_l
\label{mu11}
\end{equation}
and obtain the system of the following inequalities
\begin{equation}
\mu_i \ge \gamma_{ij} + \sum_{l=1}^n \alpha_l^{ij} a_l \quad for \ all  \ j \ne j(i), k(i).
\label{mu21}
\end{equation}
The following straightforward lemma gives a necessary condition
of permanency of the system \eqref{chem}.
\begin{lemma}
{Assume a tropically truncated system is permanent. Then, for each $i \in \{1,\ldots, n \}$, the $i$-th equation
of this system contains at least two terms. The terms should have different signs for coefficients $p_{ij}$, i.e., one term should be
a production one, while  another term should be a degradation term.}
\label{lemmaequil}
\end{lemma}
We call ``tropical equilibration'', the condition in Lemma \ref{lemmaequil}. This condition means
that permanency is acquired only if at least two terms of different signs have the maximal order,
for each equation of the system \eqref{chem}. This idea is not new, and can be traced back to
Newton.

The tropical equilibration condition can be used to determine the renormalization exponents, by the
following algorithm.

{\em Step 1}. For each $i$ let us choose a pair $(j(i), k(i))$ such that $j, i \in \{1,\ldots, m_i\}$ and
$j < k$. The sign of the corresponding terms should be different.

{\em Step 2}. We resolve the linear system of algebraic equations
\begin{equation}
\gamma_{i j(i)} - \gamma_{i k(i)}= -\sum_{l=1}^n \alpha_l^{ij(i)} a_l  + \sum_{l=1}^n \alpha_l^{ik(i)} a_l,
\label{alg}
\end{equation}
for $a_l$, together with the inequalities (\ref{mu21}).

Notice that although that Step 2 has polynomial complexity, the tropical equilibration problem has an exponential number of choices
at Step 1.

Assume that, as a result of this procedure, we obtain the system
\begin{equation}
\frac{d\bar x_i}{dt} =\epsilon^{\mu_i} (F_{i}^+(\bar {\bf x}) - F_{i}^-(\bar {\bf x})), \quad F_{i}^{\pm}=p_{ij^{\pm}} \bar {\bf x}^{\alpha_{\pm}^{ij}}.
\label{chemtrop1}
\end{equation}
One can expect that, in a "generic" case\footnote{supposing
that multi-indices $\vec{\alpha_{ij}}$ are chosen uniformly, by generic we understand
almost always except for cases of vanishing probability,
see also \cite{gorban-dynamic}}, all $\mu_i$ are mutually different, namely
\begin{equation}
0=\mu_1 < \mu_2 < ... < \mu_{n-1} <\mu_n.
\label{chain}
\end{equation}
We can now state a sufficient condition for permanency.
Let us consider the first equation (\ref{chemtrop1}) with $i=1$ and let us denote $y=\bar x_1, z=(\bar x_2, ..., \bar x_n)^{tr}$.
In this notation, the first equation becomes
\begin{equation}
\frac{dy}{dt} =f(y)=b_1(z) y^{\beta_1} - b_2(z) y^{\beta_2}, \quad b_1, b_2 > 0, \quad \beta_i \in {\bf R}.
\label{chemtrop}
\end{equation}
Since $\mu_2 > 0$, one has that $z(t)$  is a slow function of time and thus we can suppose that $b_i$ are constants (this step can be rendered rigorous by using the concept of invariant
manifold and methods from \cite{henry1981geometric}).
The permanency property can be then checked in an elementary way.   All rest points of (\ref{chemtrop})
are roots of $f$. If $f > 0$, $y(t)$ is an increasing time function and if $f < 0$, $y(t)$ is a decreasing time function.
A single root $y_1$ of $f$ within $(0, +\infty)$ is given by
\begin{equation}
y_1=\frac{b_2}{b_1}^{d}, \quad d=\frac{1}{\beta_1 - \beta_2}.
\label{chemR}
\end{equation}
These properties entail the following result:
\begin{lemma}{
Equation \ref{chemtrop} has the permanence property if and only if
 $$
0 < \beta_1 < \beta_2, \quad {\bf i}
$$
or
 $$
 \beta_1 < \beta_2 < 0, \quad {\bf ii}
$$
or
$$
\beta_1 < 0, \quad \beta_2 >0. \quad {\bf iii}
$$
For fixed $z$, in these cases we have
$$
y(t,z) \to y_0, \quad as \ t \to \infty.
$$
}
\label{permcrit}
\end{lemma}
The generic situation described by the conditions \eqref{chain} lead to trivial ``chain-like''
relaxation towards a point attractor, provided that we have permanency at each step. This result is the nonlinear analogue of the similar result that monomolecular networks with total separation relax as chains and can only have stable point attractors \cite{gorban-dynamic}.

The following theorem describes a less trivial situation, when limit cycles are possible.
\begin{theorem}
{Assume $0=\mu_1 < \mu_2 < ... < \mu_{n-1} \leq \mu_n$ holds. If the procedure, described above, leads to the permanency property at each step, where $i=1,2,..., n-2$,
and the last two equations have a globally attracting hyperbolic rest point or globally attracting hyperbolic limit cycle,
then the tropically truncated system is permanent and has an attractor of the same type.
Moreover, for sufficiently small $\epsilon$ the initial system also is permanent for initial data from  some
appropriate domain $W_{\epsilon, a, A}$ and has an analogous
 attracting hyperbolic rest point (limit cycle) close to the attractor of the truncated system. If the rest point (cycle) is not globally attracting,
then we can say nothing on permanency but, for sufficiently small $\epsilon$,  the initial system  still has an analogous
attracting hyperbolic rest point (limit cycle) close to the attractor of truncated system and the same topological structure.}
\label{bigtheorem}
\end{theorem}
Finally, let us note that tropical equilibrations with permanency imply the
existence of invariant manifolds. This allows to reduce the number of variables of the model while preserving good accuracy in the description of the dynamics. The following Lemma
is useful in this aspect.
\begin{lemma}{
Consider the system
\begin{equation}
\frac{dy}{dt} =f(y)=b_1(z) y^{\beta_1} - b_2(z) y^{\beta_2}, \quad b_1, b_2 > 0, \quad \beta_i \in {\bf R}.
\label{chemtrop4}
\end{equation}
\begin{equation}
\frac{dz}{dt} =\lambda F(y, z),
\label{Zdt}
\end{equation}
where $z \in {\bf R}^m$, $\lambda >0$ is a parameter
 and the function $F$ enjoys the following properties. This function lies in an H\" older class
$$
F \in C^{1+r}, \quad r > 0,
$$
and the corresponding norms are uniformly bounded  in $\Omega=(0, +\infty) \times W$, for some open domain $W \subset {\bf R}^m$:
$$
 |F|_{C^{1+r}(\Omega)} < C_2.
$$
Assume one of conditions {\bf i, ii, iii} of Lemma \ref{permcrit} holds.
We also suppose that $b_i$ are smooth functions of $z$ for all $z$ such that $|z| >\delta_0 >0$. Assume that
$z \in W$ implies $|z| > \delta_0$.

Then, for sufficiently small $\lambda < \lambda_0(C_2, b_1, b_2, \beta_1, \beta_2, \delta)$ equations (\ref{chemtrop4}), (\ref{Zeq})
have a locally invariant and locally attracting manifold
\begin{equation}
y =Y(z, \lambda),  \quad Y \in C^{1+r}(W),
\label{Zeq}
\end{equation}
and $Y$ has the asymptotics
\begin{equation}
Y(z, \lambda)=y_1(z) + \tilde Y,  \quad \tilde Y \in C^{1+r}(W),
\label{Zeq1}
\end{equation}
where
\begin{equation}
|\tilde Y(z, \lambda)|_{C^{1+r}(W)} < C_s\lambda^{s},  \quad s >0.
\label{Zeq2}
\end{equation}
}
\label{invariantmanifold}
\end{lemma}
{\bf Proof}. The proof is standard, follows from Theorems
in \cite{henry1981geometric}, Ch. 9.

\section{A paradigmatic cell cycle model and its tropicalization}
\subsection{Description of the model}
 We study here the cell cycle model proposed by Tyson \cite{tyson1991modeling}. This model
 mimics the interplay between cyclin and cyclin dependent kinase cdc2 (forming the maturation
promoting factor MPF complex) during the progression of
the cell cycle. The model demonstrates that this biochemical system can function as an oscillator, or converge to a steady state with large MPF concentration, or behave as an excitable switch. The three
regimes can be associated to early embryos rapid division,
metaphase arrest of unfertilized eggs, and growth controlled division of somatic cells, respectively. This model takes into account autocatalytic activity of MPF (positive feed-back). It can be
described as a nonlinear cycle of biochemical reactions and corresponds to the following set of
differential equations:
\begin{eqnarray}
 y_1' & =\epsilon^{-3} k_9 y_2 - \epsilon^{-6}k_8 y_1 + k_6 y_3, \, &
 y_2' = \epsilon^{-6}k_8 y_1 - \epsilon^{-3}k_9 y_2 - \epsilon^{-2}k_3 y_2 y_5, \notag \\
  y_3' &=\epsilon^{2}k_4' y_4 + \epsilon^{-2}k_4 y_4 y_3^2 - k_6 y_3, \, &
  y_4' = - \epsilon^{2}k_4' y_4 - \epsilon^{-2}k_4 y_4 y_3^2 + \epsilon^{-2}k_3 y_2 y_5, \notag \\
  y_5' &=\epsilon^{2} k_1 - \epsilon^{-2}k_3 y_2 y_5,
\label{tyson67}
\end{eqnarray}
Here $k_i>0$ are rate constants, $y_i,\, i\in [1,5]$ are concentrations of cdc2, p-cdc2 (phosphorylated kinase), cyclin-p:cdc2 complex (active MPF), cyclin-p:cdc2-p complex (inactive MPF), and cyclin, respectively.
With respect to the original model we have introduced a small parameter $\epsilon >0$
to cope with the order of the rate constants ($\epsilon = 0.1$ in the original model).

The system  \eqref{tyson67} has the conservation law
\begin{equation}
 y_1(t) +  y_2(t) +  y_3(t) +  y_4(t) = 1,
\label{conservation0}
\end{equation}
where the value $1$ (total initial concentration of kinase cdc2) was chosen by convenience.

\subsection{Tropical equilibrations and model reduction}
Let us apply the tropical equilibration principle.
To this aim, we renormalize the variables,
\begin{eqnarray}
y_i =\epsilon^{a_i} \bar y_i.
\end{eqnarray}
Let us substitute these relations into  the system of equations.  As a result, we obtain
\begin{eqnarray}
 \bar y_1' & =\epsilon^{-3+a_2-a_1} k_9 \bar y_2 - \epsilon^{-6}k_8 y_1 + k_6 \epsilon^{a_3-a_1} \bar y_3, \, &
 \bar y_2' = \epsilon^{-6+ a_1-a_2}k_8 \bar y_1 - \epsilon^{-3}k_9 \bar y_2 - \epsilon^{-2+a_5}k_3 \bar y_2 \bar y_5,  \notag \\
  \bar y_3' & =\epsilon^{2+a_4 -a_3}k_4'  \bar y_4 + \epsilon^{-2+a_3 +a_4}k_4 \bar y_4 \bar y_3^2 - k_6 \bar y_3, \, &
  \bar y_4' = - \epsilon^{2}k_4' \bar y_4 - \epsilon^{-2 +2a_3}k_4 \bar y_4 \bar y_3^2 + \epsilon^{-2+a_2 +a_5-a_4}k_3 \bar y_2 \bar y_5, \notag \\
  \bar y_5' & =\epsilon^{2-a_5} k_1 - \epsilon^{-2 + a_2} k_3 \bar y_2 \bar y_5.
  \label{tyson68}
\end{eqnarray}
The system  \eqref{tyson68} has the conservation law
\begin{equation}
\epsilon^{a_1} \bar y_1(t) + \epsilon^{a_2}\bar y_2(t) + \epsilon^{a_3}\bar y_3(t) + \epsilon^{a_4}\bar y_4(t)=1.
\label{conservation}
\end{equation}
In order to compute the exponents $a_i$ we use tropical equilibrations together with
the conservation law \eqref{conservation}. There are $2^4$ variants of tropical
equilibrations. To our surprise, there is only one solution for the exponents values.
We can show that all possible
equilibrations of the variables $y_3$, $y_4$ and $y_5$ uniquely set the values of two
exponents, $a_3=2$, $a_4=0$.

Let us consider the variants with
respect to the equilibrations of the variables $y_1$ and $y_2$.
Denoting by $T_i$ the $i^{th}$ term in
the equation, we have  the following situations:

1) In eq. for $\bar y_1$: $T1=T2, \ T3 <= T1$,
In eq. for $\bar y_2$: $T1=T2,  \ T3 <= T1$.

2) In eq. for $\bar y_1$: $T1=T2, \ T3 <= T1$,
In eq. for $\bar y_2$: $T1=T3,  \ T2 <= T3$.

3) In eq. for $\bar y_1$: $T2=T3, \ T1 <= T2$,
In eq. for $\bar y_2$: $T1=T2,  \ T3 <= T1$.

4) In eq. for $\bar y_1$: $T2=T3, \ T1 <= T2$,
In eq. for $\bar y_2$: $T1=T3,  \ T2 <= T1$.

In the variant 1 (Case {\bf I}) the tropical equilibrations do not fix the values of the exponent
$a_5$ and we get

\begin{equation}
 a_1=7 - a_5, \quad a_2=4 - a_5, \quad a_3=2, \ a_4=0,  \ a_5 \geq -1.
 \label{KGB}
\end{equation}
In the variants 2,3,4 (Case {\bf II}) the exponents are uniquely determined from equilibrations
and we obtain
\begin{equation}
 a_1=8, \quad a_2=5, \quad a_3=2, \ a_4=0,  \ a_5=-1.
 \label{KGB1}
\end{equation}
However,  \eqref{KGB1} and the conservation law \eqref{conservation} are incompatible, therefore,
the case {\bf II} can be rejected. In the case {\bf I} the conservation law takes the form
$
\epsilon^{7-a_5} \bar y_1(t) + \epsilon^{4- a_5}\bar y_2(t)  + \bar y_4(t)=1+o(1),
$
as $\epsilon \to 0$. Assuming that $\bar y_2(0)=O(1)$ and $\bar y_4 \ne 1$ (it is reasonable, since it is a "generic case"), we obtain $a_5=4$. Thus, the only possible situation is variant 1 (Case {\bf I}) and the corresponding set of exponents is:
\begin{equation}
 a_1=3, \quad a_2=0, \quad a_3=2, \ a_4=0,  \ a_5 =4.
 \label{KGB2}
\end{equation}
Let us note that the terms $T1$ and $T2$ in the equations for
the variables $y_1,y_2$ correspond to direct and reverse rates
of a phosphorylation/dephosphorylation cycle transforming
$y_1$ into $y_2$ and back. Thus, biochemically, (Case {\bf I})
corresponds to the quasi-equilibrium of this cycle.
Furthermore, the equilibration of all the variables leads to the exponents
\eqref{KGB}. In this case, $2+a_4 -a_3 = -2+a_3 +a_4 =0$,
$2 = -2 + 2a_3 = -2 + a_2 + a_5 - a_4$, meaning
that the tropical equilibrations of the variables $y_3$, $y_4$
are triple (in each equation, all three terms have the same order).

We finally obtain the following renormalized system
\begin{eqnarray}
 \bar y_1' & =\epsilon^{-6} (k_9 \bar y_2 - k_8 \bar y_1) + k_6 \epsilon^{-1} \bar y_3, \, &
 \bar y_2' = \epsilon^{- 3}(k_8 \bar y_1 - k_9 \bar y_2) - \epsilon^{2}k_3 \bar y_2 \bar y_5,  \notag \\
 \bar y_3' & =k_4' \bar y_4 + k_4 \bar y_4 y_3^2 - k_6 \bar y_3, \, &
  \bar y_4' =  \epsilon^{2}(-k_4' \bar y_4 - k_4 \bar y_4 \bar y_3^2 + k_3 \bar y_2 \bar y_5), \notag \\
  \bar y_5' & =\epsilon^{-2} (k_1 -  k_3 \bar y_2 \bar y_5).
  \label{tyson69}
\end{eqnarray}
The structure of the system \eqref{tyson69} emphasizes the multiple time
scales of the model. The fastest variables are in order $y_1$,
then $y_2$ and $y_5$. The variables $y_3$ and $y_4$ are slow.

Assume that
  \begin{equation}
  \bar y_2 > \delta >0.
  \label{y2d}
\end{equation}
This important assumption ensures the existence of an invariant manifold and
will be justified, a posteriori.

Then, from the last equation (\ref{tyson69})  we obtain the relation
$$
\bar y_2 \bar y_5= k_1/k_3 + O(\epsilon^2),
$$
which represents the equation of an invariant manifold.

In turn, this relation
leads to the following equations for $\bar y_3, \bar y_4$
\begin{eqnarray}
  \bar y_3' & =k_4' \bar y_4 + k_4 \bar y_4 \bar y_3^2 - k_6 \bar y_3, \, &
  \bar y_4' =  \epsilon^{2}(-k_4' \bar y_4 - k_4 \bar y_4 \bar y_3^2 + k_1).
  \label{y3y4}
\end{eqnarray}
A second invariant manifold equation is defined by the equation
\begin{equation}
\bar y_1=k_8^{-1}(k_9 \bar y_2 + k_6 \epsilon^{5}\bar y_3).
\label{y11}
\end{equation}
 Remind that $\bar  y_3$ is a slow variable. Then, for $\bar y_2$ we have
 \begin{equation}
\bar y_2'=\epsilon^2( k_6 \bar y_3 - k_1).
\label{y12}
\end{equation}
System \eqref{y3y4} represents a two-dimensional reduced model of the initial
five-dimensional system. This result shows that tropical equilibrations can
be used for model reduction.

The solutions of (\ref{y3y4}) either tend to the stable equilibrium
\begin{equation}
 \bar y_4=\frac{k_1}{k_4' +  k_4 (k_1/k_6)^2}, \quad \bar y_3=k_1/k_6,
  \label{man33}
\end{equation}
or, if this equilibrium is unstable, to a limit cycle.

Based on the general Theorem \ref{bigtheorem} we can assert the following:
\begin{theorem}{Assume \eqref{y2d} holds with $\delta > 0$.
If the shorted system (\ref{y3y4}) has a stable hyperbolic limit cycle, then, under above conditions, for sufficiently small
$\epsilon$ the five component system (\ref{tyson67}) also has a stable limit cycle.
If the shorted system (\ref{y3y4}) has a stable hyperbolic equilibrium, then, under above conditions, for sufficiently small
$\epsilon$ the five component system (\ref{tyson67}) also has a stable hyperbolic equilibrium}.
\end{theorem}
We have studied the system (\ref{y3y4}) analytically and numerically.
The numerical simulations confirm the criteria of cycle existence
both for small $\epsilon$ and for $\epsilon=O(1)$. For small epsilon
the cycle has a singular structure. The amplitude of $\bar y_3$ and the cycle period increase in $\epsilon$, approximatively, as $\epsilon^{-2}$ (the assertion about period is natural since the rate of $\bar y_4$ is $O(\epsilon^2)$).

Hyperbolicity can be straightforwardly checked for the rest point \eqref{man33}, by computing the eigenvalues of the linearized system. Denote by $Y=(\bar y_3, \bar y_4)$. When the rest point
$Y^0=(\bar y_3^0, \bar y_4^0)$ is hyperbolic and stable we have
 the following estimate
\begin{equation}
 |Y(t) - Y^0| < C_1 \exp(-c_1\epsilon^2 t)
  \label{est}
\end{equation}
with some $C_1, c_1 >0$ holds.    Integrating \eqref{y12}
for $\bar y_2$ over interval $[0, \tau]$ gives
\begin{equation}
|\bar y_2(\tau) - \bar y_2(0)| < \epsilon^2 C_1 c_1^{-1}  =o(1)
\label{y2below}
\end{equation}
uniformly in $\tau > 0$ as $\epsilon \to 0$. This yields that $\bar y_2(t) > \delta$ if $\bar y_2(0) > 2\delta$
and therefore, $\bar y_2(t)$ does not go to zero for large $t$, justifying the estimate
\eqref{y2d} needed for the existence of an invariant manifold.
The case of a limit cycle is discussed in the next subsection.

\subsection{Singular limit cycle and hybrid dynamics}
Up to this point, the tropical ideas were used for reducing the dynamics of the model.
In this section we show that the tropicalization heuristic is well adapted for decomposing
the limit cycle into slow and fast modes, providing a hybrid description of the dynamics.

Let us note that in a hybrid, excitable system, it is possible that not all variables
are equilibrated. Also, the system can have more than two different equilibrations and associated
invariant manifolds, and jump from one invariant manifold to another during the dynamics.
Let us consider that the variables $y_1,y_2,y_5$ are equilibrated as above, but now, only one among the variables $y_3$ or $y_4$ are equilibrated. We have four situations:

1) In eq. for $\bar y_3$: $T1=T3, \ T2 <= T1$, 2) In eq. for $\bar y_3$: $T2=T3, \ T1 <= T2$,

3) In eq. for $\bar y_4$: $T1=T3, \ T2 <= T1$, 4) In eq. for $\bar y_4$: $T2=T3, \ T1 <= T2$.

Combined with the conservation law condition \eqref{conservation}, variants 1 and 2
lead to the same triple tropical equilibration as before (Case {\bf I}) and exponents
\eqref{KGB2}. We denote the corresponding invariant manifold $\mathcal M_1$. The renormalized equations are the same as \eqref{y3y4}.

Variant 3 can be rejected by the general permanency criterion given by Lemma \ref{permcrit}.
Variant 4 (Case {\bf III}) satisfies the permanency criterion and leads to
\begin{equation}
 a_1=3, \quad a_2=0, \quad a_3 = 0, \quad a_4=4,  \quad a_5=4.
 \label{KGB3}
\end{equation}
This corresponds to a double equilibration (two equal terms)
of the variable $y_4$, the variable $y_3$ being not equilibrated.
We denote the corresponding invariant manifold $\mathcal M_2$. The
renormalized equations read
\begin{eqnarray}
  \tilde y_3' & =\epsilon^{6} k_4' \tilde y_4 + \epsilon^{2} k_4 \tilde y_4 \tilde y_3^2 - k_6 \tilde y_3, \, &
  \tilde y_4' =  - \epsilon^{2} k_4' \tilde y_4 + \epsilon^{-2} ( - k_4 \tilde y_4 \tilde y_3^2 + k_1).
  \label{y3y4bis}
\end{eqnarray}

\begin{figure}
\begin{centering}
\includegraphics[width=80mm]{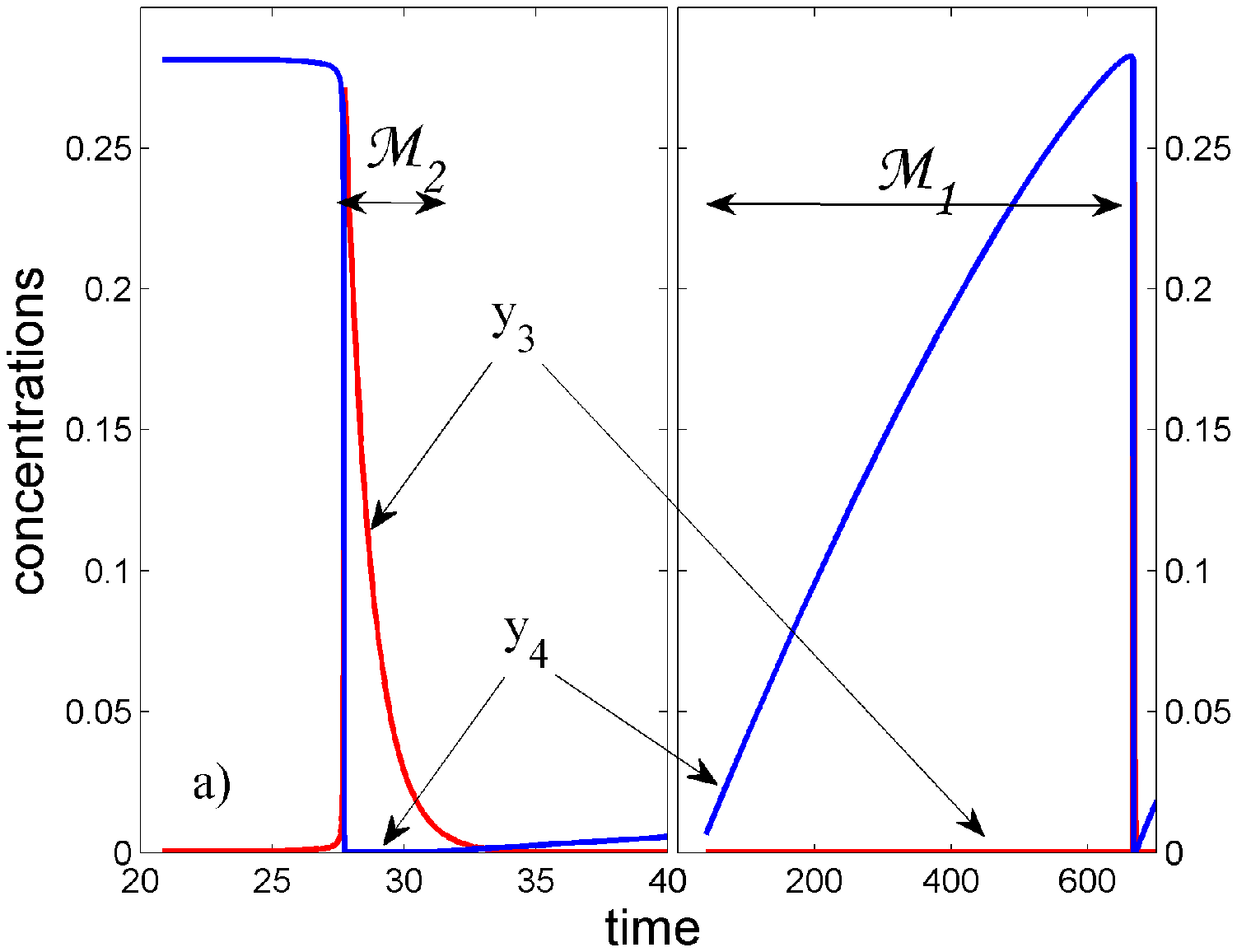}
\includegraphics[width=80mm]{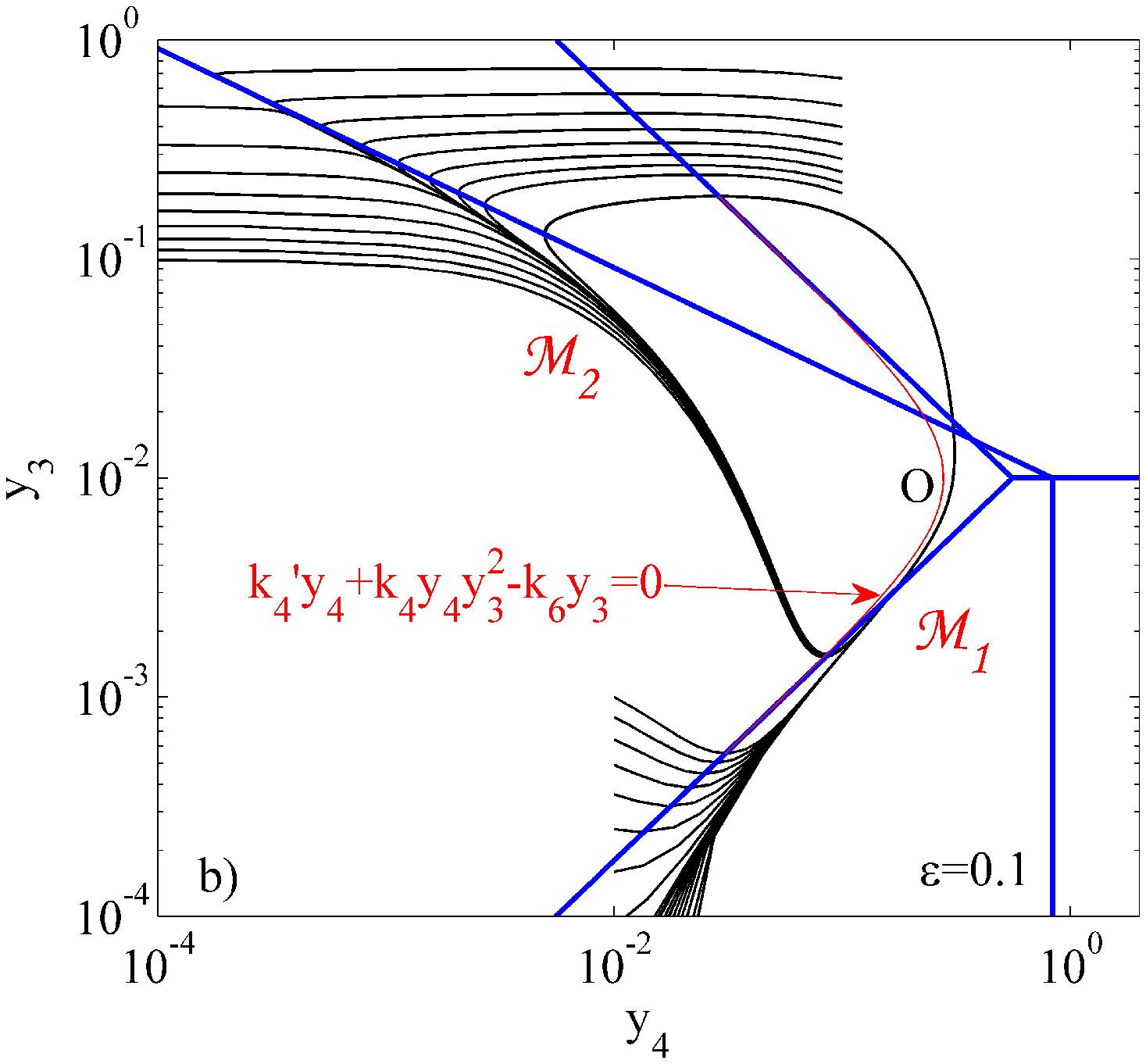}
\includegraphics[width=80mm]{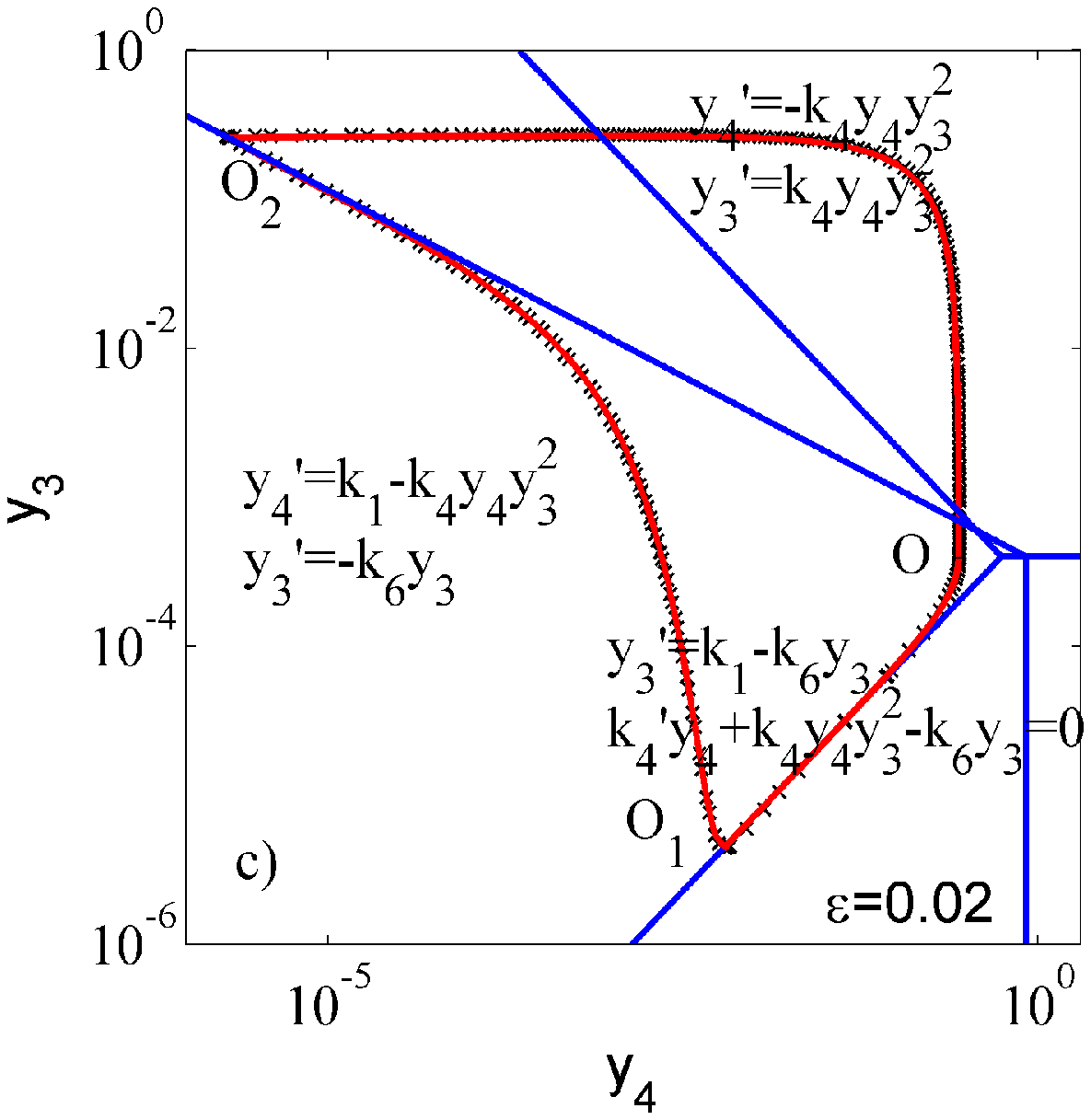}
\end{centering}
\caption{Limit cycle behavior of the paradigmatic cell cycle model from \cite{tyson1991modeling}. (a) The three main processes during the embryonic cell cycle are, in order of the timescales, fast increase of $y_3$ (active MPF, triggering mitosis), slower decrease of $y_3$  and very slow increase of $y_4$ (inactive MPF). (b) Two invariant manifolds corresponding to the two slow processes are close to the tropical manifolds (blue lines) and result from equilibration of the variables (equilibration of $y_4$ corresponds to $\mathcal M_2$ and equilibration of $y_3$ corresponds to $\mathcal M_1$). (c) A three modes hybrid approximation of the cell cycle (in red) compared to the original limit cycle (black crosses).
\label{fig1}
}
\end{figure}

We can provide a hybrid description of the cell cycle, by decomposing the periodic orbit
into three modes (Fig.\ref{fig1}). The first mode is the slowest and has the longest
duration. It consists in the dynamics on the slow invariant
manifold $\mathcal M_1$ at low values of $y_3$, and can be described by the
algebraic-differential system
$\tilde y_3'  = \epsilon^2 k_1 - k_6 \tilde y_3,\,
 k_4' \tilde y_4 + k_4 \tilde y_4' \tilde y_3^2 - k_6 \tilde y_3=0$
 (part between $O_1$ and $O$ of the orbit in Fig\ref{fig1}c)).
 The exit from the invariant manifold $\mathcal M_1$ occurs at
 a critical point (point $O$).
 The next slowest mode corresponds to the decrease of $y_3$
 (part between $O_2$ and $O_1$ of the orbit in Fig\ref{fig1}c))
 and can be described by two terms truncated system
 $\tilde y_4'  =\epsilon^{-2}( k_1 - k_4 \tilde y_4 \tilde y_3^2),\,
 \tilde y_3'= - k_6 \tilde y_3$. Finally, there is a fast mode, corresponding
 to the fast increase of $\tilde y_3$ (part
 between $O$ and $O_2$ of the orbit in Fig\ref{fig1}c)) and described
 by the truncated system  $\tilde y_4'  =  - \epsilon^2 k_4 \tilde y_4 \tilde y_3^2,\,
 \tilde y_3'= \epsilon^{-2} k_4 \tilde y_4 \tilde y_3^2$.
One can notice (Fig\ref{fig1}c)) that this hybrid approximation
is very accurate for small $\epsilon$. At a distance from the tropical
manifold, the hybrid orbit coincides with the one generated by the two
terms or by the complete tropicalization. However, close to the tropical
manifold, the two term and the complete tropicalization are less
accurate than the hybrid approximation described above.

Below we state rigorous estimates describing  the slow movement on $\mathcal M_1$
and  the fast jump towards $\mathcal M_2$.
The two terms description of the dynamics on $\mathcal M_2$ is a direct consequence of
the Proposition~\ref{comparison} and Lemmas~\ref{permcrit},\ref{invariantmanifold}.

To simplify notation, we rewrite the system \eqref{y3y4} for $y_3, y_4$ as
\begin{eqnarray}
x' & = y + y x^2  - k_0 x,
\label{xeq} \\
y' & = \epsilon^2( -y - y x^2  + k_1).
\label{yeq}
\end{eqnarray}
We can obtain such a presentation by a linear variable change.

Let us define the functions
$$
X(y)=\frac{k_0^2 - \sqrt{k_0^2 - 4y^2}}{2y},
$$
$$
X_{+}(y)=\frac{k_0^2 + \sqrt{k_0^2 - 4y^2}}{2y},
$$
and the points
$$
y_0=k_0/2, \quad x_0=X(y_0)=1.
$$
\begin{lemma}
{Solutions of (\ref{xeq}), (\ref{yeq}) with initial data $x(0), y(0)$ such that
\begin{equation}
\label{indat}
0 < \delta_0 < x(0) <  X_{+}(y_0) -\delta_0, \quad  0 < y(0) < y_0-\delta_0,
\end{equation}
where $\delta_0$ is a small positive number independent on $\epsilon$,
satisfy
\begin{equation}
\label{attr}
|x(t) -X(y(t))| < C_1 (\epsilon + \exp(-c_1 \delta_0 t)), \quad t < T_0(x(0), y(0), \epsilon)
\end{equation}
this estimate holds while
\begin{equation}
\label{T0}
 y(t) < y_0-\delta_2, \quad \delta_2 >0.
\end{equation}
}
\end{lemma}
Let us find some estimates of solutions at $y=y_0, x=x_0$. Our goal is to prove that at this point $x(t)$ starts
to increase sharply. After this, the terms $\pm yx^2$ play the main role in the equations \eqref{xeq},\eqref{yeq}, and
the other terms can be removed while the $x$-component is big.

Let us introduce new variables $u, v$ by
$$
x-x_0=v, \quad u=y-y_0.
$$
For $u, v$ one obtains
\begin{equation}
v'=\frac{k_0}{2} v^2 + u(1+ (1+v)^2),
\label{eqv}
\end{equation}
\begin{equation}
u'=\epsilon^2 ( - (\frac{k_0}{2}+u)(1 + (1+v)^2)+ k_1))=\epsilon^2 g(u,v).
\label{equ1}
\end{equation}
Let us consider for this system the Cauchy problem with initial data
\begin{equation}
v(0)=v_0, \quad  u(0)=u_0.
\label{eqin}
\end{equation}
\begin{lemma}
{Consider the Cauchy problem (\ref{eqv}), (\ref{equ1}) and (\ref{eqin})
under assumptions that
$$
u_0 =\kappa > 0, \quad |v_0| <\delta_5,
$$
and
\begin{equation}
k_1 > k_0,
\label{kk}
\end{equation}
where $\delta_0, \delta_5$ are small enough (but independent on $\epsilon$). 
Let $A$ be a large positive number independent of $\epsilon$.
Then within some interval 
$t \in (\tau_0(\delta_5, k_0, k_1, A), \tau_1(\delta_5, k_0, k_1, A))$
one has
\begin{equation}
u(t) > 0,  \quad v(t) \ge A, \quad v(\tau_0)=A.
\label{grow}
\end{equation}
}
\end{lemma}
This result can be reinforced. Actually, $x(t)$ attains values of the order $O(\epsilon^{-2})$.
\begin{lemma}
{Assume
\begin{equation}
x(t_1)=A >> 1,  \quad y(t_1) \ge k_0/2.
\label{indx}
\end{equation}
Then
\begin{equation}
x(t) \ge (A^{-1} - \frac{1}{2}\sigma t)^{-1},
\label{indxx}
\end{equation}
\begin{equation}
y(t)  \ge \sigma,
\label{indxy}
\end{equation}
for $\sigma, A$ such that
\begin{equation}
\sigma  A > 2k_0, \quad \sigma > k_0\exp(-2\sigma^{-1}), 
\label{cond}
\end{equation}
and $t$ such that
\begin{equation}
(A^{-1} - \frac{1}{2}\sigma t) \ge \epsilon^{2}.
\label{condT}
\end{equation}
}
\end{lemma}

\section{Conclusion}
We showed that tropical ideas can be usefully
employed to reduce and hybridize polynomial or rational
dynamical systems occurring in modelling
the molecular machinery of the cell cycle.
The main idea consists in keeping only the dominant monomial terms in the right hand side of
the ordinary differential equations. Depending on the position in phase space, one should keep one,
two, or more such terms. The places where two or more monomial terms are equal form the so-called
tropical manifolds. The one term approximation is valid far from the tropical manifolds, whereas
close to tropical manifolds several dominating terms of opposite signs can equilibrate each
other. These ``tropical equilibrations'' of the dominating terms slow down the
dynamics and produce attractive invariant manifolds.

The possible applications of this method are multiple. Generally, the method can be
used to obtain simplified models. In the example studied here, we have started with a
five variables model, that has been reduced to two variables and hybridized.
The modes of the hybrid model have the simple structure of monomial differential
or differential-algebraic
equations. Two general methods that we called complete and two terms tropicalizations
provide description of the modes and of the mode changes.
However, these general procedures may lead to inaccurate approximations when the
full model does not satisfy permanency globally. In such cases,  more thorough
analysis is needed. We have shown that the model of embryonic
cell cycle has essentially three modes with different timescales,
namely slow accumulation of cyclin, rapid activation of MPF and intermediately
rapid degradation of cyclin and inactivation of MPF. The fastest mode is described by
monomial ODEs, whereas the less fast modes correspond to tropical equilibrations
and are described by differential-algebraic equations.


Several improvements and developments are needed in order to apply these methods
at a larger scale. The computation of tropical equilibrations suffers
from combinatorial explosion. However, for the biochemical network used as
working example, the number of solutions seems to be very small compared
to the large combinatorics of monomial terms. There is hope, that
once formulated in constraint logic programming, the problem of equilibrations
could be efficiently computed in practice as a constraint satisfaction problem.
Also, effective methods are needed to compute the transitions between
modes.
The main difficulty here is related to walls (segments of
the tropical manifolds) crossing. Near walls, two or more terms are dominant.
When these terms are equilibrated,
orbits remain close to the walls and are contained in invariant manifolds.
The complete or two terms tropicalizations provide general heuristic for
mode transitions. These approximations mail fail close to walls.
For instance, as we showed in a previous paper \cite{SASB2011},
the complete tropicalization predicts sliding modes that evolve on the wall
and stay thus close to orbits of the full system. However, these sliding modes
can be too long, leaving the wall when the orbits of the full system are
already far away.
In order to get an accurate description of the behavior near
such walls we had to compute invariant manifolds.
Although this is generally much simpler than integrating the full set
of equations, it could become difficult for tropical equilibrations
involving more than two terms. Future work will be dedicated to
developing general methods for this problem.

{\small
\bibliography{hybridbis,Bioinfo,radulescu}
\bibliographystyle{eptcs}
}

\section{Appendix: proofs}

{\bf Proof of Lemma 3.2}
Let us consider the equation
\begin{equation}
x'= y(0) + y(0) x^2  - k_0 x=f_0(x).
\label{xeq0}
\end{equation}
One observes that
$$
f(x) < -\delta_4 < 0,  \quad x \in (X(y(0))+\delta_0, X_{+}(y_0)-\delta_0),
$$
$$
f(x) > \delta_4, \quad x \in (\delta_0, X(y_0)-\delta_0),
$$
and
$$
f'(x)\vert_{x=X(y_0)} < -\delta_5 < 0.
$$
Therefore,
if $0 < x(0) < \frac{k_0^2 +\sqrt{k_0^2 - 4y(0)^2}}{2y(0)} -\delta_0$, then
the solution $x(t)$ attains a small $\delta$- neighborhood of $X(y(0))$ within a bounded time interval
$T_1(\delta, \delta_0, \delta_4)$:
$$
|x(t, x_0, y_0) - X(y(0))| < \delta, \quad t=T_1(\delta, \delta_0, \delta_4).
$$
Within a small $\delta$ - neighborhood of $X(y(0)$ we set $u=x-X(y(0))$ and then we can rewrite (\ref{xeq0}) as follows :
\begin{equation}
u'=  -\kappa u + h(u), \quad |h(u)| < C_1 u^2, \quad u(T_1)=\delta.
\label{xeq0N}
\end{equation}
where $\kappa > 0$ is independent of $\delta$. Then, if $\delta$ is small enough, we have
that $u(t) < \delta$ for all $t > T_1$ and
$$
|u(t)| < C(T_0)\exp(-\kappa t/2), \quad t > T_0.
$$
Let us compare now the solution $x(t)$ of (\ref{xeq0}) and the corresponding solution $\bar x(t, x_0, y_0)$ of
(\ref{xeq}) with the same initial data.
For $x(t)- x(0)=w$ one has, since $y(t) - y(0) < C(T_0) \epsilon^2$ on any bounded interval $t \in [0, T_0]$,
$$
w_t = a(t) w  + \epsilon^2 g(x, t, w), 0 < t \le T_0.
$$
with a smooth function $g$, $a(t)$ is bounded function. Now the Gronwall inequality implies
$$
|w(t)| < C_2(T_0) \epsilon^2, \quad t \in [0, T_0].
$$
Therefore, one has
$$
|\bar x(t, x_0, y_0) - X(y(t))| < \delta, \quad t=T_2(\delta, \delta_0, \delta_4).
$$
Since the function $X(y)$ defines a smooth, locally attracting (for $y < y_0$) invariant manifold,
 this proves our assertion. One can prove this in another, elementary way. Let us define
 $u=\bar x - X(y)$. Then
\begin{equation}
u'=  -\kappa(t) u + h(u) + O(\epsilon), \quad |h(u)| < C_1 u^2, \quad u(T_1)=\delta,
\label{xeq0Neps}
\end{equation}
where $\kappa(t) > \kappa_0 $ while $y < y_0-\delta_0$.  Again one has $u(t) < 2\delta$ for $t > T_1$ (while $y <y_0-\delta_0$).
Thus,
$$
u' \le -\frac{\kappa_0}{2} u + O(\epsilon)
$$
that entails the need estimate \ref{attr}.

{\bf Proof of Lemma 3.3}
Let us consider the Cauchy problem
\begin{equation}
w'=\frac{k_0}{2} w^2,  \quad w(0)=v_0=v(0)
\label{eqw}
\end{equation}
It is clear, by the comparison principle, that
$$
v(t) \ge w(t)
$$
while $u(t) >0$. Consequently,  the assertion $w(\tau_0) =A$ proves  the lemma.
Let us prove first that if $\delta_6$ is small enough (but independent on $\epsilon$), for some $t=\tau_1$ we have
\begin{equation}
v(\tau_1)=\delta_6.
\label{eqw}
\end{equation}
Without loss of generality, we assume that
\begin{equation}
v(t)< \delta_6, \quad 0 < t < \tau_1, \quad v(\tau_1)=\delta_6.
\label{eqw2}
\end{equation}
Let $\delta_6>0$ be small enough such that
\begin{equation}
 - \frac{k_0}{2}(1 + (1+\delta_6)^2)+ k_1 >0.
\label{equ}
\end{equation}
Such a choice of $\delta_6$ is possible due to (\ref{kk}).

Assume $u(t)\ge 0$ within some time interval $[0, \tau_2]$,  and $u(\tau_2)=0$ for some $\tau_2 < \tau_1$. Since $u(0)=0$, we have $\tau_2 >0$.
We can suppose without loss of generality that $\tau_2$ is the first moment, where $u(\tau_2)=0$.
Then
$$
u'(\tau_2) \le 0.
$$
But then we obtain a contradiction with (\ref{equ}) at $t=\tau_2$, since the right hand side of this equation is positive at this time moment.

Therefore, we have shown that $u(t) >0$ for all $t$ from $[0, \tau_1]$ if $v(t) < \delta_6$ for such $t$.
Then $v(t) \ge w(t)$ on this time interval.
The function $w$ can be found, and an easy computation gives
\begin{equation}
w(t)=(w^{-1}(0)- 2k_0^{-1}t)^{-1})=(u_0 - 2k_0^{-1} t).
\label{WW}
\end{equation}
Assume that $v(t) < \delta_6$ for all $t$. Then (\ref{WW}) holds for all $t$, but $w(\tau_1) > \delta_6$ for some $\tau_1$.
We have obtained a contradiction, thus (\ref{eqw}) is proved.

Let us prove that $v(t)=A$.
Let us consider an interval $[t_1, T]$ such that
\begin{equation}
|v(t)| < A, \quad t \in [t_1, T], \quad v(t_1)=\delta_6.
\label{int}
\end{equation}
Then (\ref{equ}) implies
\begin{equation}
u' \le \epsilon^2 ( (\frac{k_0}{2}+|u|)(1 + (1+A)^2)+ k_1)),
\label{equ5}
\end{equation}
that gives, by the Gronwall lemma,
\begin{equation}
u(t) < C_2 u(0) \exp(C_1(A) \epsilon^2 t).
\label{equ6}
\end{equation}
Within interval $(t_1, T]$ one has then
\begin{equation}
u(t) < C_4 \kappa, \quad t \in [0, T].
\label{equ6}
\end{equation}
and, therefore,
\begin{equation}
v' \ge \frac{k_0}{2} v^2 - C_4 \kappa, \quad t \in [0, T].
\label{eqv6}
\end{equation}
Suppose that
$$
C_4 \kappa  < \frac{k_0}{8} \delta_6.
$$
Then (\ref{eqv6}) entails
\begin{equation}
v' \ge \frac{k_0}{4} v,  \quad t \in [0, T],
\label{eqv7}
\end{equation}
This gives
$$
v(t) \ge \delta_6 \exp( \frac{k_0}{4} (t- t_1)).
$$
This leads to a contradiction for $T$ large enough (however, let us remark that $T$ is uniform in $\epsilon)$.

{\bf Proof of Lemma 3.4}

{\em Remark}: to satisfy (\ref{cond}) it suffices to set $\sigma =2k_0A^{-1}$ with a large $A$.

Suppose that either the  estimate (\ref{indxx}) (case A) or the second estimate (\ref{indxy}) (case B)
is violated at some $T$ but the both inequalities
 hold for all $t_1 \le t < T$.
We can set $t_1=0$.
Let us consider the case A. Since (\ref{indxy}) hold, one has
$$
x' \ge \sigma x^2, \quad x(0)=A, \quad t \in [0, T].
$$
This implies
$$
x'/x^2=(-1/x)' \ge \sigma, \ x(0)=A.
$$
Thus,
$$
x \ge (A^{-1} -\sigma t)^{-1}.
$$
For $t=T$ this result gives a contradiction with (\ref{indxx}).

Let us consider the case B. Since (\ref{indxx}) hold, one has
$$
y'(t) \le  -\epsilon^2 (y (A^{-1} -\sigma t)^{-1} -k_1), \quad y(0)=k_0/2, \quad t \in [0, T].
$$
This implies
$$
y(t) \le \frac{k_0}{2} \exp( -\epsilon^2 \int_0^t (1 + (A^{-1} -\sigma s)^{-2}) ds) +
\epsilon^2 k_1 \int_0^t \exp( -\epsilon^2 \int_\tau^t (1 + (A^{-1} -\sigma s)^{-2}) ds) d\tau.
$$
for $t \in [0, T]$.
Notice
$$
\int_0\tau^t (1 + (A^{-1} -\sigma s)^{-2}) ds =t-\tau +  \sigma^{-1} ((A^{-1} -\sigma t)^{-1}-
(A^{-1} -\sigma \tau)^{-1}).
$$
Thus, for sufficiently small $\epsilon$, taking into account (\ref{condT}) one has
$
y(T) \le \frac{k_0} \exp(- 2/\sigma).
$
Under condition (\ref{cond}) this result gives a contradiction with (\ref{indxy}).

{\bf Proof of Theorem 2.7}

{\bf i}  Suppose that the tropically truncated system (TTS) has a globally attracting compact invariant set $\mathcal A$. Let be $\Pi$ be an open neighborhood
of this set. We can choose this neighborhood as a box that contains $\mathcal A$. Then, for all initial data
$x(0)$, the corresponding trajectory $x(t), x(0)$ lies in $\Pi$ for all $t > T_0(x_0, \Pi)$.  Therefore, our TTS is permanent.
Here we do not use that the cycle (rest point) is hyperbolic.

{\bf ii} Permanency of the initial system follows from hyperbolicity of $\mathcal A$. Hyperbolic sets are persistent
(structurally stable) (Ruelle 1989). Since this set is globally attracting, all TTS is structurally stable (as a dynamical system).
This implies that the initial system has a  hyperbolic attractor close to $\mathcal A$, since initial system is a small perturbation
 of TTC in $\Pi$.

{\bf iii} If the set $\mathcal A$ is only locally attracting, the last assertion of Theorem follows from persistency of hyperbolic sets.

\end{document}